# Incommensurate Spiral Spin Order in $CaMn_2Bi_2$ observed via High Pressure Neutron Diffraction

Madalynn Marshall[1‡], Haozhe Wang[2‡], Antonio M. dos Santos[1], Bianca Haberl[1], Weiwei Xie[2*]

1. Neutron Scattering Division, Neutron Sciences Directorate, Oak Ridge National Laboratory, Oak Ridge, Tennessee 37831
2. Department of Chemistry, Michigan State University, East Lansing, Michigan 48824

[‡] M.M. and H.W. contributed equally. * Email: xieweiwe@msu.edu

## Abstract

High pressure neutron diffraction is employed to investigate the magnetic behavior of $CaMn_2Bi_2$ in extreme conditions. In contrast to antiferromagnetic ordering on Mn atoms reported at ambient pressure, our results reveal that at high pressure, incommensurate spiral spin order emerges due to the interplay between magnetism on the Mn atoms and strong spin-orbit coupling on the Bi atoms: sinusoidal spin order is observed at pressures as high as 7.4 GPa. Competing antiferromagnetic order is observed at different temperatures in the partially frustrated lattice. This research provides a unique toolbox for conducting experimental magnetic and spin dynamics studies on magnetic quantum materials via high pressure neutron diffraction.

# Introduction

Neutron scattering is a particularly powerful technique for studying materials because it is relatively sensitive to common elements like oxygen, because a thermal neutron's energy is similar to that of some important processes in materials, and because neutrons have a magnetic moment, making them sensitive to magnetic structure. The downside of using neutrons is that they interact only weakly with matter, and so relatively large sample volumes are needed. This especially hampers high pressure characterization of materials by neutron scattering because the volumes required are often much too high for experiments to be performed. In this study we succeed in performing our high-pressure neutron study of $CaMn_2Bi_2$, in which we observe a change in the magnetic ordering present by performing neutron powder diffraction (NPD) experiments at high pressures, due to an increase in the influence of spin-orbit coupling (SOC) according to our calculations.

SOC is the interaction between the intrinsic spin of electrons and their orbital motion in an electric field. This phenomenon is observed in heavy metals and commonly leads to energy level splitting, resulting in strong and adjustable magnetic ordering and enhanced magnetocrystalline anisotropy.[1-3] The presence of strong SOC in magnetically active metals can also cause the emergence of new electronic and magnetic phases,[4-6] such as topological insulator CaAgX (X = P, As)[7] and Weyl semi-metals TaAs[8,9] and $Co_3Sn_2S_2$[10,11], with implications for quantum information processing and spintronics.[12,13] Most studies of magnetic materials with strong SOC have focused on 4$d$ and 5$d$ electron-based transition metals like those based on Ru or Ir, 4$f$ electron-based rare-earth metals, and Bi containing compounds where the influence of SOC should not be a surprise.[14,15]

In contrast to materials with only SOC on a single type of atoms, this work addresses the possibility that magnetism and strong SOC can also result in unconventional behavior when present on different kinds of atoms within a single compound. One illustration of this is the magnetic topological insulator $MnBi_2Te_4$, which is based on ferromagnetic layers of MnTe interleaved with those of the topological insulator $Bi_2Te_3$.[16] In the case of $CaMn_2Bi_2$, however, it has been reported that the semiconducting behavior observed stems from hybridization between the Mn 3$d$ states and Bi 5$p$ states, rendering it a potential hybridization gap semiconductor.[17] Nevertheless, under ambient pressure, the antiferromagnetic order on the Mn sublattice in $CaMn_2Bi_2$ is impervious to the strong SOC on neighboring Bi atoms. As a result, it is of interest to determine whether modulating the interactions between the Mn and Bi atoms in $CaMn_2Bi_2$ via applied pressure will induce novel magnetic and electronic phenomena. Neutron diffraction, interpreted via first-principles calculations, is the method employed here to investigate this possibility.

To investigate the magnetic and electronic behavior of $CaMn_2Bi_2$, we utilized high-pressure neutron diffraction techniques. A previous crystal structure determination conducted using high-pressure single crystal X-ray diffraction, sensitive to the structure of a material, and requiring samples several orders of magnitude smaller in volume than those employed here, revealed the emergence of a new monoclinic phase of $CaMn_2Bi_2$ at pressures exceeding 2.35 GPa.[18] This resulted in a transition from a low-pressure phase based on a puckered honeycomb Mn sublattice to one based on quasi-one-dimensional (quasi-1D) zigzag Mn chains, leading to the possibility that geometric magnetic frustration may be present. Magnetic frustration can result in incommensurate spin ordering, including helical, cycloidal, and sinusoidal wave magnetism.[19-23] This phenomenon has been observed in systems such as $MnSi$[24] and $MnP$[25], where the transition associated with an incommensurate magnetic structure and suppression of long-range magnetic order with pressure has led to quantum states such as superconductivity.

Thus, here we report, determined by utilizing high-pressure neutron diffraction techniques, our observation that high pressure induces sinusoidal wave spin ordering in $CaMn_2Bi_2$. Combining our experimental findings with first-principles calculations, we have identified the essential interplay of SOC and hybridization for the observed magnetic responses in this system. Our results provide a unique example of materials where the incommensurate spin order is induced by pressure-tuning the interactions between 3$d$ electron-based magnetism (on the Mn atoms) and strong 5$p$-electron-based SOC (on the Bi atoms).

## Experimental

**Materials Synthesis.** For single crystals of $CaMn_2Bi_2$, the Bi self-flux method was employed by mixing Ca, Mn, and Bi with a molar ratio of 1:2:10 in an alumina crucible sealed in a quartz tube. The tube was heated to 1000 °C at a rate of 180 °C/h and held there for 48 hours. After that, the sample was slowly cooled down to 400 °C at a rate of 6 °C/h. Hexagonal-shape single crystals (~2×2×2 $mm^3$) of $CaMn_2Bi_2$ were obtained after removing excess Bi by centrifuging. The $CaMn_2Bi_2$ single crystals were crushed and then ground well for high-pressure neutron powder diffraction experiments.

**High Pressure Neutron Diffraction.** The high-pressure neutron diffraction experiments on $CaMn_2Bi_2$ were conducted on the SNAP high-pressure diffractometer at the Spallation Neutron Source at the Oak Ridge National Laboratory. The detectors were positioned at 65° and 90° with respect to the direct beam, enabling access to a *d*-spacing range of up to about 8 Å. To conduct these experiments, the powder sample, of ~1 g was loaded into a null-scattering Ti-Zr alloy encapsulated gasket, and pressure was applied using a Paris-Edinburgh (PE) VX5 press fitted with cubic BN single toroidal anvils. The press was installed at the beamline with the load axis in the vertical direction. The sample temperature was controlled by liquid nitrogen flow in copper channels closely attached to both anvils. Pressure was incrementally increased in steps of 0.1–0.2 GPa, followed by data collection, to a maximum pressure of 7.4 GPa. Data reduction was performed using the Mantid package (Workbench version 6.2), and Rietveld refinements were conducted using FullProf Suite.[26-28] The refined parameters for $CaMn_2Bi_2$ consisted of lattice parameters, atomic coordinates, and thermal displacements.

## Results and Discussion

**Quasi-one-dimensional (quasi-1D) Frustrated Mn Lattice:** In our previous high pressure single crystal X-ray diffraction (XRD) measurements, $CaMn_2Bi_2$ undergoes a plane-to-chain structure phase transition from puckered Mn−Mn honeycomb layer to quasi-1D Mn−Mn chains at $P$ ~2.4 GPa accompanied by a large volume collapse.[18] A high precision structure determination was provided before, thus the high pressure crystal structure evolution was used for the calibration of the high-pressure NPD measurements here: the pressure was back-calculated and normalized to the ambient-pressure value determined from the unit cell volume of the sample. Temperature dependent NPD measurements were performed in a series of temperatures 85, 100, 120, 150, 180, and 290 K at pressure of ~6.9 GPa, and 85, 120, and 290 K at pressure of ~7.4 GPa. The resulting nuclear structure at high pressure was confirmed to be consistent with our previous single crystal XRD report[18] and is presented in **Figure 1**. At pressure of ~6.9 GPa, magnetic diffraction peaks were started to be observed in the 180 K dataset, with 958 magnetic Bragg peaks in the studied $d$ spacing range in total. Considering the phase diagram of Bi with pressure, both Bi-II and Bi-III phases were also included in the refinement leading to an overall improved result, although only trace Bi was observed as a remained flux in crystal growth of $CaMn_2Bi_2$.

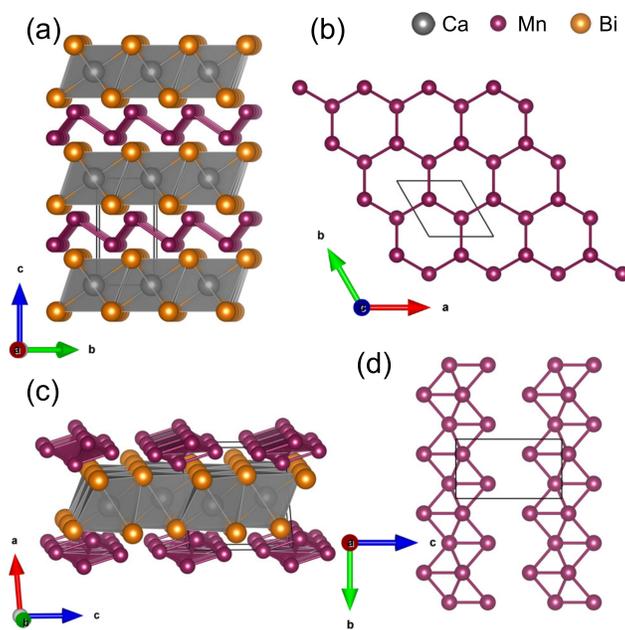

**Figure 1 (a)** Crystal structure of $CaMn_2Bi_2$ under ambient pressure[18] showing Ca@$Bi_6$ octahedra. **(b)** View of Mn puckered honeycomb sublattice. **(c)** Crystal structure under ~7 GPa consisting with distorted edge-shared Ca@$Bi_6$ octahedra. **(d)** View of Mn quasi-1D chain sublattice.

The crystal structure of monoclinic $CaMn_2Bi_2$ at high pressures is characterized by quasi-1D chains of Mn atoms. Structurally, quasi-1D chains can be classified into two categories, namely, "zigzag" and "ladder" chains. The term, zigzag chains, is used to describe adjacent chains that are shifted by 1/2 of the chain repeat distance, whereas ladder chains have aligned adjacent chains or without a shift. Quasi-1D structural chains provide a potentially unique platform for observing incommensurate spin ordering, including helimagnetic structures, observed in several ternary chalcogenides and halides, depending on the intra and interchain interactions.[34-39]

**Determination of the Incommensurate Propagation Vector of the Magnetic System:** To obtain a suitable magnetic spiral propagation vector, a Gaussian function was fitted to the prominent magnetic structure peaks at ~4.0 Å and ~5.0 Å to determine the magnetic peak widths. A doublet and a triplet were observed at ~4.0 Å and ~5.0 Å, respectively, and a $k$-vector (0, 1/2, 1/8) was proposed as the magnetic scattering periodicity by the FullProf Suite through using the $k$-search functionality. However, refinement with $k = (0, 1/2, 1/8)$ resulted in several incorrect peak positions, shown in **Figure 2**. As such, a refined $k$ vector $k' = (0, \pm\delta, \pm\gamma)$ was tested, which yielded lower $\chi$ and $R$ factor values by fully accounting for the observed magnetic peak positions as well as for their widths. The resulting incommensurate magnetic propagation vector, of (0, 0.48, 0.13), is similar but very significantly different from the commensurate value of $k = (0, 1/2, 1/8)$. The results obtained from the refinement are shown in **Table 1**. **Table 2** shows that $k'$ deviates by a small amount, though further from $k = (0, 1/2, 1/8)$ with increasing pressure.

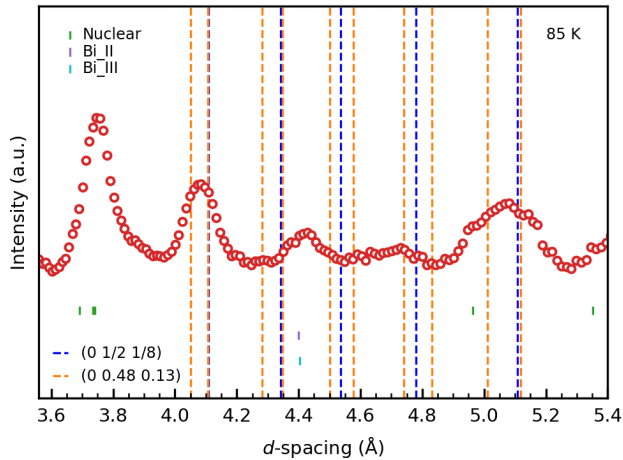

**Figure 2** The NPD pattern of CaMn$_2$Bi$_2$ at 85 K and ~7.4 GPa comparing $k$ vectors $k = (0, 1/2, 1/8)$ marked in blue and $k' = (0, \pm\delta, \pm\gamma)$ marked in yellow.

**Table 1** The agreement factors for the magnetic structure refinements at 85 K. Magnetic $R$-factor, $\chi$, $R_p$, $R_{wp}$, and $R_{exp}$ refinement parameters from neutron powder diffraction refinements for $k$ vectors $k = (0, 1/2, 1/8)$ and $k' = (0, \pm\delta, \pm\gamma)$ at 85 K for ~6.9 and ~7.4 GPa.

| Pressure | k-vector | Magnetic R-factor | $\chi$ | $R_p$ | $R_{wp}$ | $R_{exp}$ |
|---|---|---|---|---|---|---|
| ~6.9 GPa | (0, 1/2, 1/8) | 11.4 | 7.50 | 0.868 | 1.53 | 0.56 |

|  | (0, 0.490, 0.130) | 10.7 | 6.49 | 0.816 | 1.44 | 0.57 |
| ~7.4 GPa | (0, 1/2, 1/8) | 15.7 | 4.00 | 0.840 | 1.26 | 0.63 |
|  | (0, 0.484, 0.137) | 14.2 | 2.80 | 0.722 | 1.07 | 0.64 |

**Table 2** The magnetic spiral propagation vectors at 85 K. Refined $k$ vectors at 85 K for ~6.9 and ~7.4 GPa.

| Pressure | Temperature | Propagation vector | | |
|---|---|---|---|---|
|  |  | $k_x$ | $k_y$ | $k_z$ |
| ~6.9 GPa | 85 K | 0 | 0.490(2) | 0.130(1) |
| ~7.4 GPa | 85 K | 0 | 0.484(1) | 0.137(1) |

**Figure 3** and **Figure S1** present the refined NPD patterns obtained at approximately 6.9 GPa at the temperature range of 85–180 K. At 180 K, the magnetic reflections are too weak to be observed or indexed. Further cooling down to 85 K, additional magnetic reflections are observed associated with the incommensurate spiral spin order in monoclinic CaMn$_2$Bi$_2$, as highlighted in **Figure 3**. These magnetic reflections can be indexed with the propagation vector for $k' = (0, \pm\delta, \pm\gamma)$. The magnetic peaks gradually increase in intensity as the temperature decreases from 180 K to 85 K. Similar behavior is observed at ~7.4 GPa (**Figure 4**).

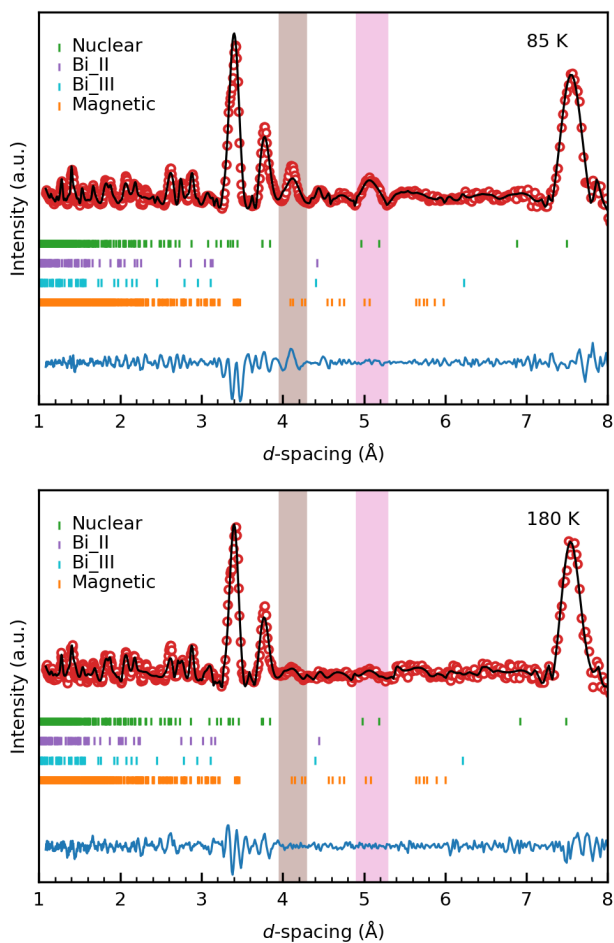

**Figure 3** The NPD refinements at ~6.9 GPa at temperatures of 85 K (upper panel) and 180 K (lower panel), where the red dotted line is the NPD pattern modelled with the black line Rietveld fitting and the corresponding residual pattern shown as the blue line. The theoretical Bragg peak positions are indicated for the $CaMn_2Bi_2$ nuclear peaks (green), Bi II (purple), Bi III (teal) and the $CaMn_2Bi_2$ magnetic peaks (orange). The magnetic peak regions are highlighted in brown and pink for visual purposes.

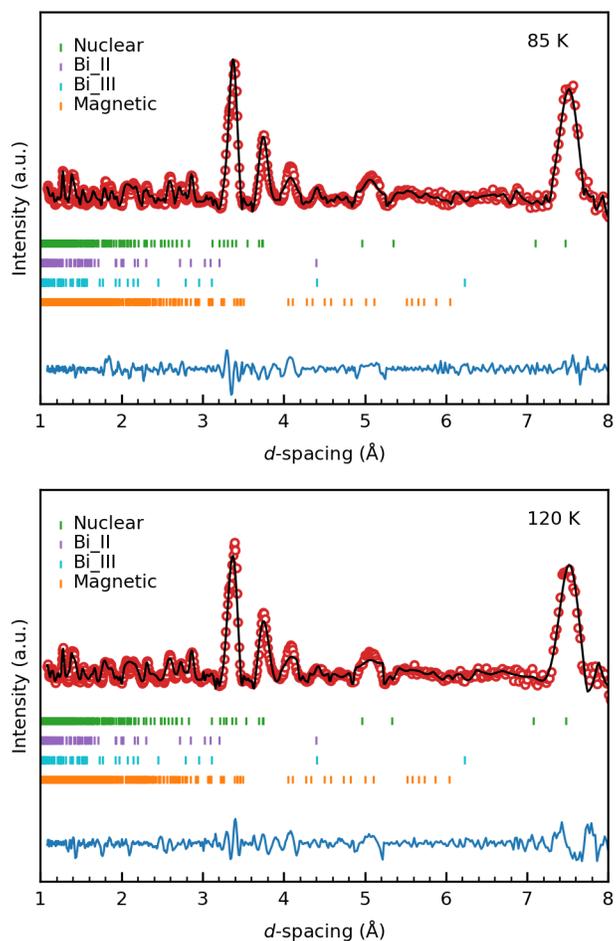

**Figure 4** The NPD refinements at ~7.4 GPa at temperatures of 85 K (upper panel) and 120 K (lower panel), where the red dotted line is the NPD pattern modelled with the Rietveld fitting (black line) and the corresponding residual pattern (blue line). The theoretical Bragg peak positions are indicated for the $CaMn_2Bi_2$ nuclear peaks (green), Bi II (purple), Bi III (teal) and the $CaMn_2Bi_2$ magnetic peaks (orange).

**Incommensurate Spiral Spin Order:** High-pressure CaMn$_2$Bi$_2$ possesses a zigzag double chain, as shown in **Figure 5**. The zigzag chain displays helimagnetic ordering. To investigate the helimagnetic ordering of high-pressure CaMn$_2$Bi$_2$, a real-space description of multi-axial helical structures with an elliptic envelope was employed using the FullProf Suite software. However, this resulted in an inadequate model as multiple magnetic peaks were not accounted for. To address this issue, additional magnetic structures were explored using the SARAh software to analyze the propagation vector $k = (0, 1/2, 1/8)$ in the parent space group $P2_1/m$. Four irreducible representations, each consisting of three basis vectors, were identified (**Table 3**). A similar result was obtained through representational analysis of the propagation vector $k' = (0, \pm\delta, \pm\gamma)$. All irreducible representations were necessary to construct an appropriate model for antiferromagnetic ordering considering the non-zero intensity of the (111) magnetic Bragg peak.

**Table 3** The irreducible representations (IR) and the corresponding basis vectors (BV) for the magnetic Mn atoms at the coordinates MnA1 = $(x, 1/4, z)$, MnA2 = $(1-x, 3/4, 1-z)$, MnB1 = $(u, 1/4, v)$ and MnB2 = $(1-u, 3/4, 1-v)$ with the associated real magnetic components in the $a$, $b$, and $c$ axis directions.

| IR | $\Gamma_1$ | | | $\Gamma_2$ | | |
|---|---|---|---|---|---|---|
| BV | $m_a$ | $m_b$ | $m_c$ | $m_a$ | $m_b$ | $m_c$ |
| MnA1 $\psi_1$ | 1 | 0 | 0 | --- | --- | --- |
| MnA1 $\psi_2$ | 0 | 0 | 1 | --- | --- | --- |
| MnA1 $\psi_3$ | --- | --- | --- | 0 | 1 | 0 |
| MnA2 $\psi_1$ | 0 | 1 | 0 | --- | --- | --- |
| MnA2 $\psi_2$ | --- | --- | --- | 1 | 0 | 0 |
| MnA2 $\psi_3$ | --- | --- | --- | 0 | 0 | 1 |
| MnB1 $\psi_1$ | 1 | 0 | 0 | --- | --- | --- |
| MnB1 $\psi_2$ | 0 | 0 | 1 | --- | --- | --- |
| MnB1 $\psi_3$ | --- | --- | --- | 0 | 1 | 0 |
| MnB2 $\psi_1$ | 0 | 1 | 0 | --- | --- | --- |
| MnB2 $\psi_2$ | --- | --- | --- | 1 | 0 | 0 |
| MnB2 $\psi_3$ | --- | --- | --- | 0 | 0 | 1 |

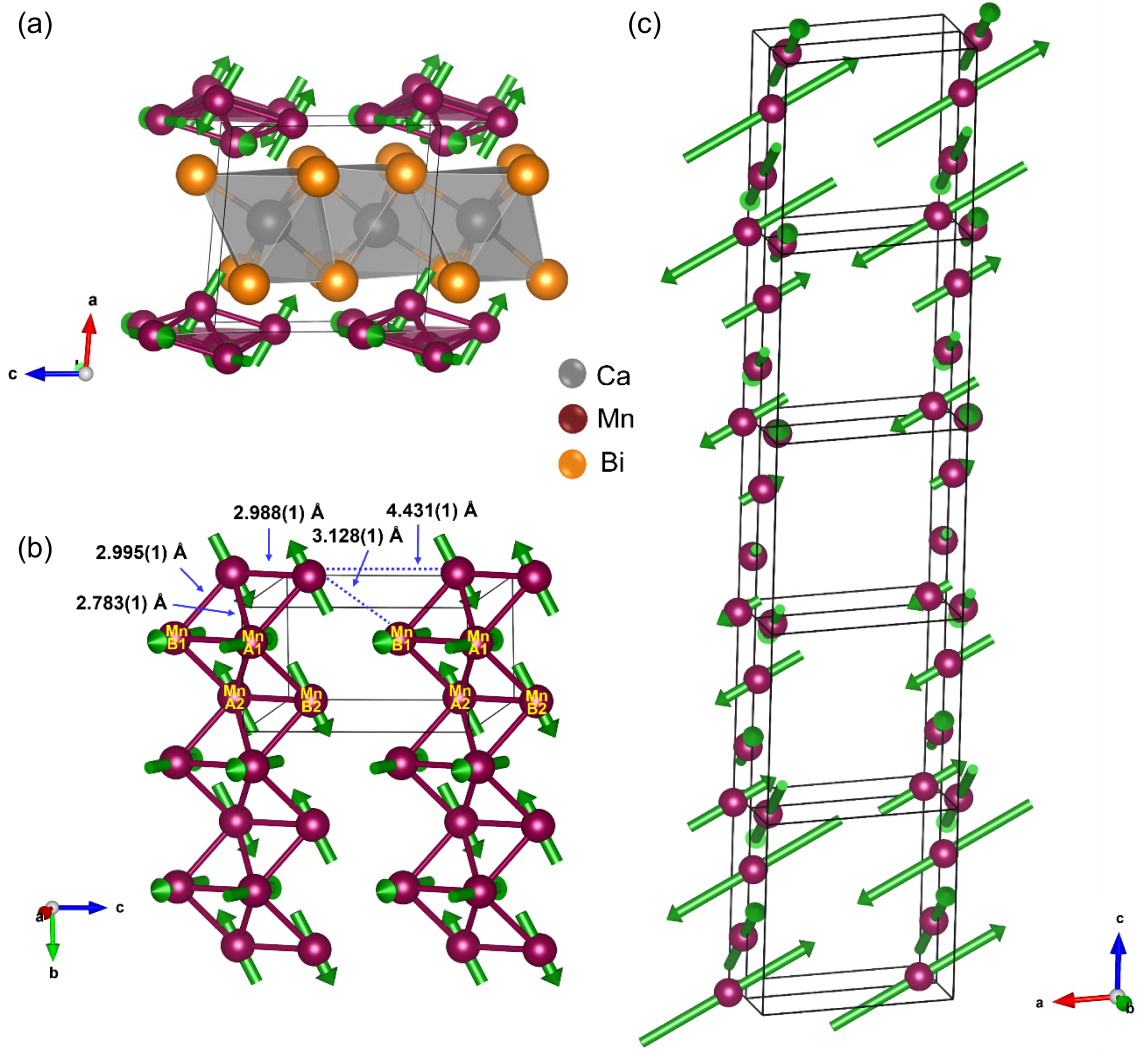

**Figure 5** The magnetic structure of CaMn$_2$Bi$_2$ at high pressure, showing **(a)** the Ca@Bi$_6$ octahedra and the quasi-1D zigzag Mn chains in the *ac* plane, **(b)** the atomic distances and labeled coordinates of the Mn quasi-one-dimensional zigzag chains along the *b* axis at 85 K and ~7.4 GPa, and **(c)** the sinusoidal wave spin ordering of the Mn moments along the *c* axis. **(a)** and **(b)** are depicted without sinusoidal ordering for visual purposes.

High-pressure monoclinic CaMn$_2$Bi$_2$ has two independent manganese atom sites, MnA and MnB. The representational analysis performed reveals that each crystallographic Mn atom site splits into two magnetic sites, leading to the coordinates: MnA1 = ($x$, 1/4, $z$), MnA2 = (1–$x$, 3/4, 1–$z$), MnB1 = ($u$, 1/4, $w$), and MnB2 = (1–$u$, 3/4, 1–$w$). Based on the symmetry of the propagation vector, magnetic structures for high-pressure CaMn$_2$Bi$_2$ can be determined by varying the coefficients $c_i$ within the FullProf Suite program for a given combination of irreducible representations. Importantly, all possible irreducible representations and combinations, i.e., $\Gamma_1$, $\Gamma_2$ and $\Gamma_1$&$\Gamma_2$, lead to sinusoidal magnetic ordering. A sinusoidal Mn spin structure has been observed in several manganese oxides, including RMnO$_3$ (R = rare earth) and the Mn zigzag chain material SrMnGe$_2$O$_6$.[23, 40, 41] The same zigzag Mn chain geometry found in SrMnGe$_2$O$_6$ is also exhibited by high-pressure CaMn$_2$Bi$_2$, while the interchain Mn-Mn distances in high-pressure CaMn$_2$Bi$_2$, 2.783(1) Å, 2.988(1) Å, and 2.995(1) Å, are much smaller than those in SrMnGe$_2$O$_6$ (3.257(3) Å), making interchain interactions a stronger source of competing spin interactions. Other zigzag Mn chain sinusoidal magnetic structures have been discovered in the triple perovskite Mn$_3$MnNb$_2$O$_9$ and the multiferroic material MnWO$_4$.[42, 43] Although, generally, at higher temperatures the sinusoidal spin structure dominates, at lower temperatures, new magnetic ordering may occur as is the case for SrMnGe$_2$O$_6$, which transitions from a sinusoidal to a cycloidal spin structure with decreasing temperatures. It may therefore be interesting to investigate the magnetic ordering for high-pressure CaMn$_2$Bi$_2$ at lower temperatures, especially considering the fact that an additional low temperature transition (labelled as $T_2$) was seen in the high-pressure resistivity measurements. Antiferromagnetic (AFM) models with a collinear arrangement of the MnA1, MnA2, MnB1 and MnB2 magnetic moments resulted in a relatively high magnetic $R$ factor and $\chi$. After allowing all coefficients $c_i$ to freely refine, a preferred AFM model was revealed consisting of colinear antiparallel MnA1 and MnB1 moments perpendicularly aligned with colinear antiparallel MnA2 and MnB2 moments. A visualization of this final magnetic structure can be seen in **Figure 5**. The $\Gamma_1$&$\Gamma_2$ combination of irreducible representations enhanced the agreement factor most. This results in a spin canting towards the $c$ axis experienced by all the Mn magnetic moments. We note that due to the complexity and low symmetry of the monoclinic structure for high-pressure CaMn$_2$Bi$_2$, a stable refinement could only be achieved when several constraints were imposed, namely MnA1 $m_a$ = MnB1 $m_a$, MnA1 $m_c$ = MnB1 $m_c$, MnA2 $m_b$ = MnB2 $m_b$ and MnA2 $m_c$ = MnB2 $m_c$. Perpendicularly aligned competing AFM ordering between Mn magnetic moments has been observed before in Mn$_2$MnReO$_6$, where the perpendicular nature was explained as a way to maximize the antisymmetric Dzyaloshinskii–Moriya interactions. However, whether the same is the case for high-pressure CaMn$_2$Bi$_2$ remains to be determined; the details of the magnetic ordering while not perfectly defined, all point to the same conclusion, namely that sinusoidal magnetic ordering within the zigzag Mn chains dominates the system.

As listed in **Table 4**, the observed magnetic moments expose the relative strengths of the potential AFM ordering schemes. At temperatures > 120 K at ~6.9 GPa the MnA2 and MnB2 moments appear to dominate, resulting in a partial frustration of the MnA1 and MnB1 moments, while at ≤ 120 K the MnA1 and MnB1 moments appear to saturate at ~4.6 $\mu_B$, which is close to the ideal value of 5.9 $\mu_B$ for Mn$^{2+}$, partially frustrating the MnA2 and MnB2 moments at ~4.0 $\mu_B$. A similar occurrence appears at ~7.4 GPa as well. Finally, the Mn magnetic moments diminish with increasing pressure in agreement with our basic model that applied pressure induces stronger hybridization of the atomic valence orbitals.

**Table 4** The total and projected onto the lattice vectors *a*, *b*, and *c* crystal axis components of the magnetic moments of Mn at ~6.9 GPa at temperatures 85, 100, 120, 150 and 180 K.

| Translation | Crystal Axis | 85 K | 100 K | 120 K | 150 K | 180 K |
|---|---|---|---|---|---|---|
| | | | **MnA1** | | | |
| | *a* | -3.20(2) | -3.20(1) | -3.19(1) | -2.0(2) | -2.18(1) |
| (0, 0, 0) | *b* | 0 | 0 | 0 | 0 | 0 |
| | *c* | 3.25(1) | 3.25(1) | 3.24(1) | 3.24(1) | 2.22(1) |
| **Total Moment** | | 4.66(2) | 4.65(1) | 4.63(1) | 3.86(2) | 3.17(1) |
| | | | **MnA2** | | | |
| | *a* | 0 | 0 | 0 | 0 | 0 |
| (0, 0, 0) | *b* | -3.23(1) | -3.23(1) | -3.22(1) | -3.23(1) | -2.21(1) |
| | *c* | -2.40(1) | -2.40(1) | -2.39(1) | -2.39(1) | -2.39(1) |
| **Total Moment** | | 4.02(1) | 4.02(1) | 4.02(1) | 4.02(1) | 3.26(1) |
| | | | **MnB1** | | | |
| | *a* | -3.20(2) | 3.20(1) | 3.19(1) | 2.0(2) | 2.18(1) |
| (0, 0, 0) | *b* | 0 | 0 | 0 | 0 | 0 |
| | *c* | 3.25(1) | -3.25(1) | -3.24(1) | -3.24(1) | -2.22(1) |
| **Total Moment** | | 4.66(2) | 4.65(1) | 4.63(1) | 3.86(2) | 3.17(1) |
| | | | **MnB2** | | | |
| | *a* | 0 | 0 | 0 | 0 | 0 |
| (0, 0, 0) | *b* | 3.23(1) | 3.23(1) | 3.22(1) | 3.23(1) | 2.21(1) |
| | *c* | 2.40(1) | 2.40(1) | 2.39(1) | 2.39(1) | 2.39(1) |
| **Total Moment** | | 4.02(1) | 4.02(1) | 4.02(1) | 4.02(1) | 3.26(1) |

## Conclusion

A comprehensive investigation of the magnetic structure of monoclinic high-pressure $CaMn_2Bi_2$ performed via high-pressure neutron powder diffraction reveals the emergence of sinusoidal spiral spin order by pressures of ~6.9 GPa. A distinctive noncollinear arrangement is observed within the zigzag chains of the magnetic Mn ions. The material exhibits spiral spin magnetic ordering with a $k$ vector that is not commensurate ($k = (0, 0.484, 0.137)$ at 85 K and ~7.4 GPa) with indications of partial frustration as the temperature varies. It may be of future interest to conduct further theoretical modeling and single-crystal neutron diffraction experiments to verify this picture of $CaMn_2Bi_2$ at high pressure.

## Acknowledgments

The work was primarily supported by the U.S. DOE-BES under Contract DE-SC0023568. A portion of this research used resources at the Spallation Neutron Source, a DOE Office of Science User Facility operated by the Oak Ridge National Laboratory.

## Supporting Information

The neutron powder diffraction refinements at ~6.9 GPa at temperatures of 100 K, 120 K, and 150 K.

# Supporting Information

# Incommensurate Spiral Spin Order in CaMn$_2$Bi$_2$ observed via High Pressure Neutron Diffraction

Madalynn Marshall[1‡], Haozhe Wang[2‡], Antonio M. dos Santos[1], Bianca Haberl[1], Weiwei Xie[2*]

1. Neutron Scattering Division, Neutron Sciences Directorate, Oak Ridge National Laboratory, Oak Ridge, Tennessee 37831
2. Department of Chemistry, Michigan State University, East Lansing, Michigan 48824

[‡] M.M. and H.W. contributed equally. * Email: xieweiwe@msu.edu

# Table of Contents

**Figure S1** The neutron powder diffraction (NPD) refinements at ~6.9 GPa at temperatures of 100 K, 120 K, and 150 K, where the red dotted line is the NPD pattern modelled with the black line Rietveld fitting and the corresponding residual pattern shown as the blue line. The theoretical Bragg peak positions are indicated for the $CaMn_2Bi_2$ nuclear peaks (green), Bi II (purple), Bi III (teal) and the $CaMn_2Bi_2$ magnetic peaks (orange).

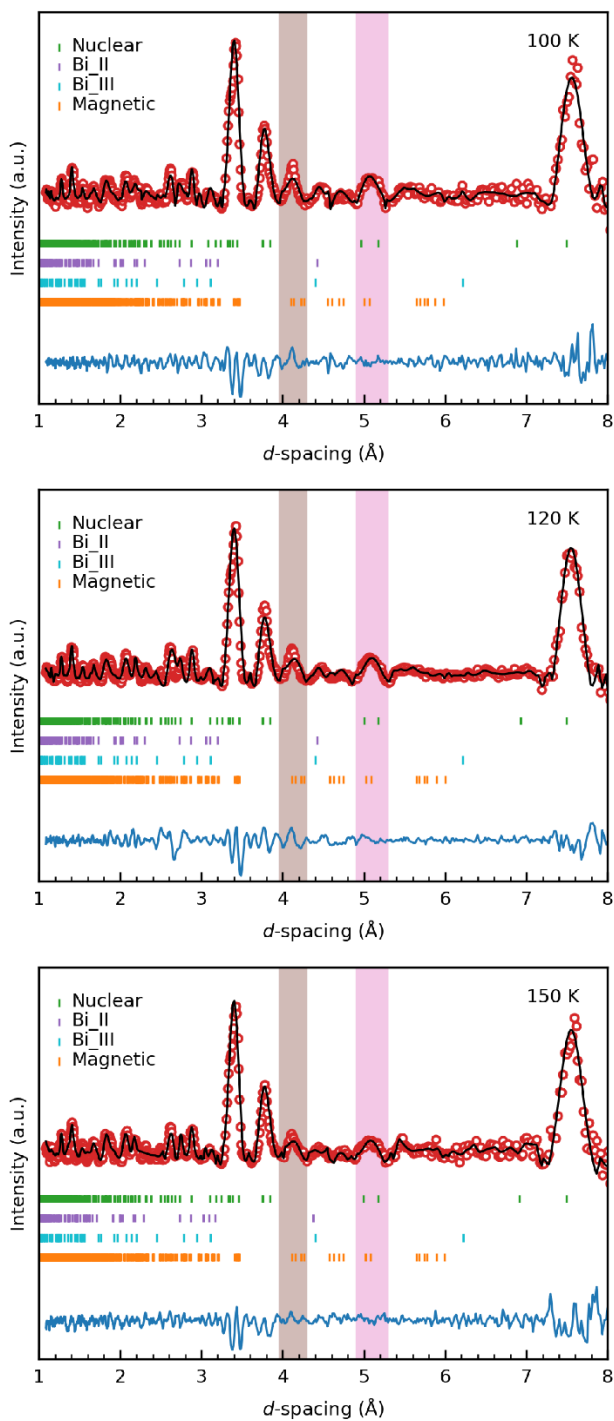